\documentclass{iopart}
 
\usepackage{epsfig}
\usepackage{here}
\usepackage{cite}

\begin{document}
   \title[]{Upper limit for the cross section
               of the overlapping scalar resonances $f_{0}(980)$
               and $a_{0}(980)$ produced in proton-proton 
               collisions in the range of the reaction threshold
   }
\author{P.~Moskal$^{1,2}$\footnote{e-mail address: p.moskal@fz-juelich.de},
        H.-H.~Adam$^3$, A.~Budzanowski$^4$, R.~Czy{\.z}ykiewicz$^2$, D.~Grzonka$^1$,
        M.~Janusz$^2$, L.~Jarczyk$^2$, B.~Kamys$^2$,
        A.~Khoukaz$^3$, K.~Kilian$^1$, C.~Kolf$^1$, P.~Kowina$^{1,5}$, T.~Lister$^3$,
        W.~Oelert$^1$, C.~Piskor-Ignatowicz$^2$, J.~Przerwa$^2$, 
        C.~Quentmeier$^3$, T.~Ro{\.z}ek$^{1,5}$,
        R.~Santo$^3$, G.~Schepers$^{1}$, T.~Sefzick$^1$,
        M.~Siemaszko$^5$,
        J.~Smyrski$^2$, A.~Strza{\l}kowski$^2$, A.~T{\"a}schner$^3$,
        P.~Winter$^1$, 
        M.~Wolke$^1$\footnote{present address: The Svedberg
                  Laboratory, Thunbergsv{\aa}gen 5A, Box 533, S--75121 Uppsala, Sweden.},
        P.~W{\"u}stner$^6$,
        W.~Zipper$^5$
      }
 
\address{ $^1$ IKP, Forschungszentrum J\"{u}lich, D-52425 J\"{u}lich, Germany}
\address{ $^2$ Institute of Physics, Jagellonian University, PL-30-059 Cracow, Poland}
\address{ $^3$ IKP, Westf\"{a}lische Wilhelms--Universit\"{a}t, D-48149 M\"unster,Germany}
\address{ $^4$ Institute of Nuclear Physics, PL-31-342 Cracow, Poland}
\address{ $^5$ Institute of Physics, University of Silesia, PL-40-007 Katowice, Poland}
\address{ $^6$ ZEL, Forschungszentrum J\"{u}lich, D-52425 J\"{u}lich, Germany}

\begin{abstract}
      Utilizing a missing mass technique we investigate the
      $pp\rightarrow ppX$ reaction scanning 
      beam energies in the range permitting to 
      create a mass close to that of the $f_0(980)$ and $a_0(980)$
      scalar resonances, but still below the $K^+K^-$ threshold
      where they decay dominantly into $\pi\pi$ and $\pi\eta$ mesons, respectively.
      Prior to the data analysis we introduce 
      a notion of the close to threshold
      total cross section for broad resonances.
      We estimated 
      for the overlapping mesons $a_0$ and $f_0$
      the total cross section to be smaller than 430~nb 
      at excess energy of Q~=~5~MeV.
      The experiment have been performed at the Cooler Synchrotron COSY using
      the COSY-11 facility.   
\end{abstract}

\pacs{13.60.Le, 13.85.Lg, 13.75.-n, 14.40.Cs, 29.20.Dh}
 
\submitto{\JPG}

\section{Introduction}

A study of the 1 GeV/c$^2$ mass range is motivated by the continuing
discussion on the nature of the scalar resonances $f_0(980)$ and $a_0(980)$,
which have been interpreted as exotic four quark states~\cite{JAFF},
conventional $q \bar q $ states~\cite{PEN,kleefeldf0}
 or molecular like $K \bar K$ bound
state ~\cite{WEIN,WANG}. \\   
Within the framework of the J\"ulich meson exchange model for
$\pi - \pi$~\cite{LOHSE} and $\pi - \eta$ scattering the $K \bar K$ interaction
dominated by vector meson exchange gives rise to a $K \bar K$ bound state
identified with the $f_0(980)$ in the isoscalar sector, while the isovector
$a_0(980)$ is concluded to be a dynamically generated threshold
effect~\cite{KREHL,haidenproc}.
Both shape and absolute scale of $\pi \pi \to K \bar K$ transitions turn out
to depend crucially on the strength of the  $K \bar K$ interaction, which,
in turn is prerequisite of a $K \bar K$ bound state interpretation of the
$f_0(980)$. 
Although the $K\bar{K}$ decay mode of $f_0(980)$ and $a_0(980)$ is rather 
weak in comparison to the dominant  $\pi \pi$ and $\pi\eta$ decay channels~\cite{PDG},
even a new theoretical analysis based on the chiral approach cannot account for the
$f_0(980)$ and $a_0(980)$ if the $K \bar K$ channel is not introduced
additionally to the $\pi \pi$ and $\pi\eta$ interaction~\cite{oller}.
An analysis of the $\pi\pi$ and $K\bar{K}$ interaction~\cite{zou} showed
that f$_{0}$ corresponds to  poles of three Rieman-Sheets which
appears physically as an object with a decay width of about 400~MeV
and a narrow peak width of about 50~MeV. The same parameters of the f$_{0}$ were found
by utilizing a unitarized quark model, according to which f$_{0}$ was interpreted as
a $q\bar{q}$ state with a large admixture of $K\bar{K}$ virtual state~\cite{tornqvist}.
The origin of the scalar resonances was also thoroughly studied 
by means of a  coupled channel  analysis considering  $\pi\pi$, $K\bar{K}$ and $\sigma\sigma$ 
meson--meson scattering~\cite{kaminski97,kaminski99}. 
Decreasing gradually the interchannel coupling
constants  it was inferred that for some solutions 
the f$_{0}$ corresponds to
the $K\bar{K}$ bound state~\cite{kaminski99,kaminski00}
at the limit of the  
fully uncoupled case.

In high energy experiments, 
the $f_{0}$ meson is observed as a resonance in the system of two pions
produced in the variety  of hadro-production 
reactions~\cite{alde94,barberis99,barberis,breakstone}
 or in  hadronic decays of heavier mesons~\cite{aitala,akhmetshin005,akhmetshin006}
or the Z$^{0}$ boson~\cite{ackerstaff,abreu}.
 These experiments study  invariant
masses of the created neutral (($\pi^{0}\pi^{0}$)~\cite{alde94,akhmetshin006,barberis} 
and charged ($\pi^{+}\pi^{-}$)~\cite{akhmetshin005,aitala,ackerstaff,breakstone,barberis99}) 
pion pairs.
Similarly,  charged~\cite{ackerstaffC5,gay} and  neutral $a_{0}$~\cite{a0_teige} mesons
were observed as a
clear signal in an invariant mass spectrum of 
the $\eta \pi$ system. 

Complementary to these approaches, which studied the interaction of
$\pi\pi$, $K\bar{K}$, and $\pi\eta$ meson pairs,
we investigate the possible
manifestation of the mesons $f_{0}$ and/or $a_{0}$ as  ``doorway states'' leading
to meson production in proton-proton collisions, namely
$pp \rightarrow p p f_{0}(a_{0}) \rightarrow p p \ Mesons $.
 By measuring the missing mass with respect to the pp-system we study the $f_{0}$ -- $a_{0}$
system as a genuine particle produced in the proton-proton collisions.
In section 3 we report of the first experimental investigation, 
which concerns the close to threshold production
of the broad neutral resonances via proton-proton collisions 
in the 1~GeV/c$^2$ mass 
range.
Moreover, we have studied the $f_{0}$ and $a_{0}$ mass range below 
the $K\bar{K}$ threshold, where they can decay 
into non-strange mesons only.

It is obvious that for the excitation of a broad resonance
the phrase ``close to threshold'' is not well defined and implies
here that the beam momentum is such that masses just in the range 
of the resonance can be excited. This issue will be discussed in section~2.

\begin{figure}[H]
\vspace{0.2cm}
\parbox{0.4\textwidth}{\epsfig{file=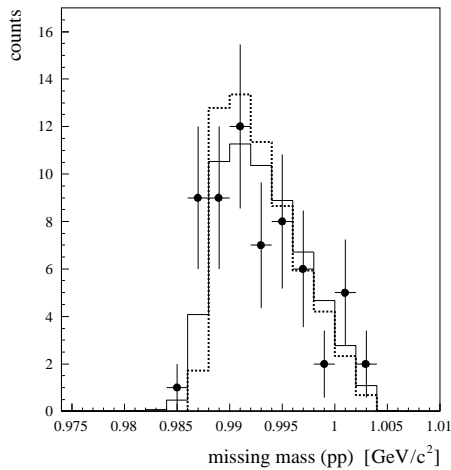,width=0.46\textwidth}}
\parbox{0.6\textwidth}{\caption{\small 
         Experimental 
         spectrum of the $K^{+}K^{-}$ invariant mass 
         measured for the reaction $pp\rightarrow ppK^{+}K^{-}$
         at a beam momentum of 3.356~GeV/c (data points).
         The width of the bins corresponds to the experimental
         resolution of the mass determination (FWHM~$\approx$~2~MeV).
         Solid and dashed lines
         show  Monte-Carlo
         simulations under the assumption of the direct 
         and resonance production, correspondingly
         \cite{quentmeierphd}.
         \label{mm_kk_f0}
        }}
\end{figure}

In a recent publication~\cite{COSY11KK} the COSY-11 collaboration presented 
data on the close-to-threshold $K^+ K^-$ production 
following the proton-proton interaction at the excess energy of Q~=~17~MeV.
The obtained distribution of the missing mass to the p--p system 
is shown in
figure~\ref{mm_kk_f0}.
Monte-Carlo simulations demonstrate that the non-resonant  $K^+ K^-$
production (solid line) is hardly distinguishable from the
resonant $pp \to pp f_0 (980) \to pp K^+ K^- $ reaction sequence 
(dashed line)~\cite{quentmeierphd}.
In fact, the
statistics of the data was not sufficient to favour one of the two processes.
The issue whether there is a chance to distinguish  between $K\bar{K}$ pairs
originating from the decay of genuine $f_{0}$/$a_{0}$ resonance
and those produced by strong $\pi\pi - K\bar{K}$ correlation is
at present under theoretical investigation\cite{haidenproc}.

 Since there exists no experimental information of the 
close to threshold total cross section for the production of the
$f_{0}$ and $a_{0}^{0}$ mesons\footnote{
Recently the ANKE collaboration reported results on the close
   to threshold production of the $a_{0}^{+}$ meson
   via the $pp \to d K^{+}\bar{K}^{0}$ reaction~\cite{vera1,vera2}},
before the presentation
of the COSY-11 measurement,
 we will estimate an expected  order of magnitude for the considered values
at the excess energy of Q~=~5~MeV at which 
theoretical predictions
for the $K^{+}K^{-}$ production  are available~\cite{bratkovskaya}.
The calculation,
based upon the one-pion exchange model and the Breit-Wigner prescription for
the $f_{0}$ resonance, 
 shows that the production of a $K\bar K$ pair through the
$f_{0}$ resonance ($pp \to pp f_0 (980) \to pp K^+ K^- $) contributes at Q~=~5~MeV,
with respect to the $K^+ K^-$ threshold,
by a factor  of at least 20 
less than the non-resonant creation.
Extrapolating the measured cross sections 
of the $pp \to pp K^+ K^- $ reaction~\cite{COSY11KK,DISTOKK}
to Q~=~5~MeV one obtains the value of 0.08~nb.
Thus combining the above information and 
additionally taking into account that the branching ratio of $f_{0}$ meson decay
into $K^+ K^-$ is about 2~$\%$~\cite{bratkovskaya},
one expects that 
very close to threshold at Q~=~5~MeV (with respect to the $ppK^{+}K^{-}$
final state) the $f_{0}$ meson should be 
produced with a cross section in the order of  0.08~nb~$\ast$~50~/~20~=~0.2~nb. 
However, the branching ratio of $f_0$ decay into $K\bar{K}$
 pairs is not well established~\cite{PDG}
and may be much smaller than the above assumed value of 2$\%$.
Naively one would expect it to be very small, since due to the energy conservation
only a few per cent of the resonance can decay into strange particles 
and moreover
only a  fraction of that part will decay into a $K^{+}K^{-}$ pair.

In case of the $NN\to NN a_{0}$ reaction the total cross section was estimated
using an effective Lagrangian approach taking into account the 
one-pion exchange mechanism and the production via t-channel
exchanges with $\pi\eta$ and $\pi f_{1}$ mesons~\cite{bratkovskaya01}.
The assumption of the positive interference between s- and u-channel
of the one-pion exchange results in the value of about~20~nb at Q~=~5~MeV,
however, an exchange of heavier mesons
was not taken into account. In addition it should be noted  that a different
 choice of  coupling constants and 
cut-off parameter $\Lambda_{\pi NN}$ would easily change the expected cross
section by a factor of five in either direction.

The above appraisement indicates that at a few~MeV above 
the production threshold
a total cross section for the $pp\rightarrow pp a_{0}(f_{0})$ reactions
is expected to be in the order of 1~nb to 100~nb. 

\section{Definition of the close-to-threshold total cross section 
         for a broad resonance}
The study of short-living (broad) particles in the $pp\rightarrow pp X$
reaction requires special care~\cite{hankudepj} when the energy in the center of mass
is close to the sum of the masses of the two protons and the average mass of the meson.
As already mentioned due to the broad mass distribution of such particles the notion of the
reaction threshold is not well defined.
It is naturally a matter of scale whether a given resonance is considered
as a broad one. Experimentally, by a broad resonance we define a particle
with its full width $\Gamma$ at half maximum of the mass distribution
(spectral function)
being much larger than the experimental accuracy of the mass determination,
or with a width such broad that the acceptance of the detection system
changes significantly over the resonance mass range. Both criteria apply for
the
measurements performed and discussed in the present contribution.\\
In the following we will propose a definition of the total cross section
as a function of the excess enery Q = $\sqrt{s} - 2 \cdot m_{p} - m_{X}$,
which is valid also when Q is small compared to the
width $\Gamma$ of the produced meson.\\

\begin{figure}[H]
\vspace{-1.2cm}
\parbox{0.47\textwidth}{\epsfig{file=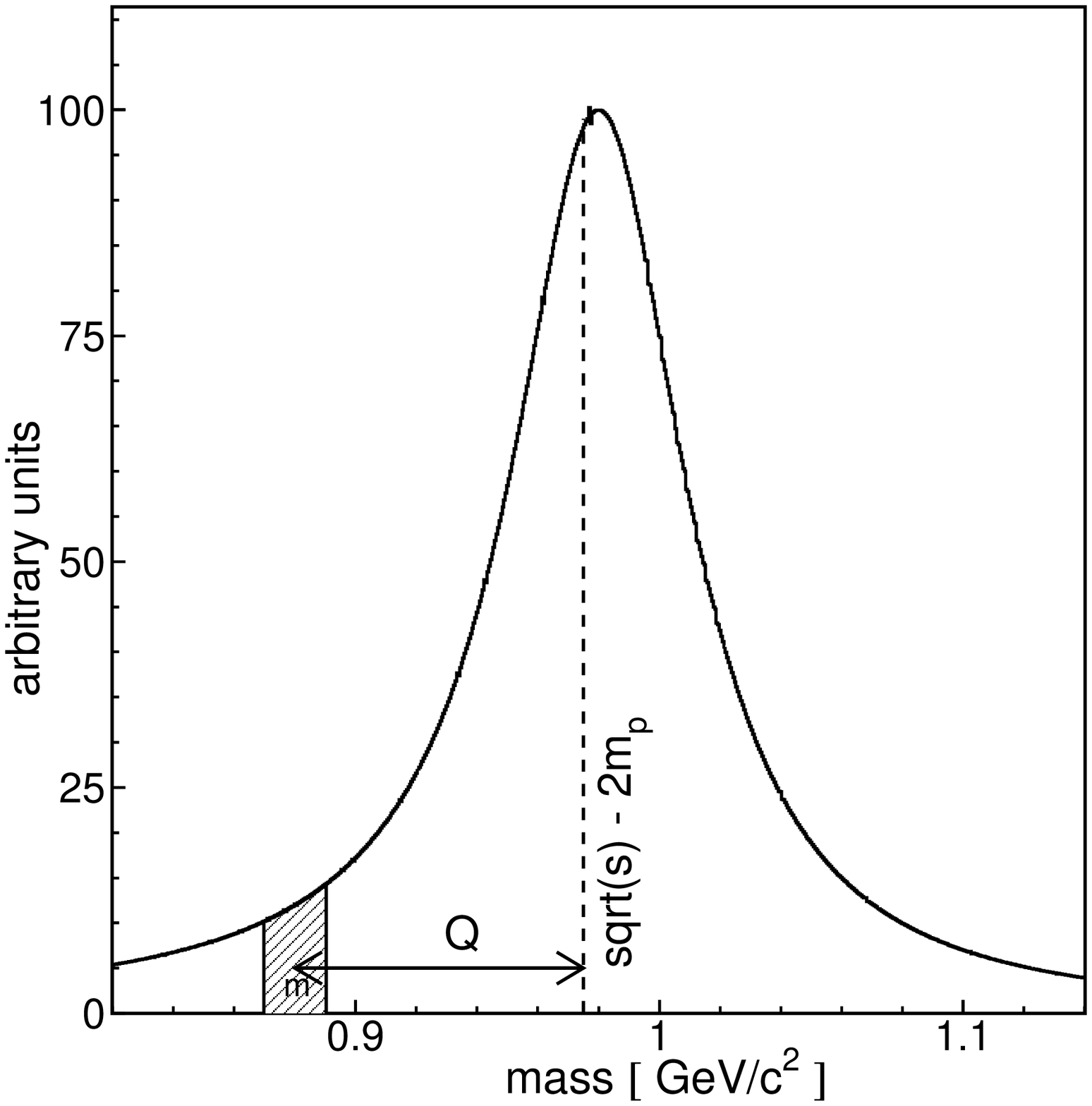,width=0.48\textwidth}}
\parbox{0.49\textwidth}{\vspace{0.6cm}\epsfig{file=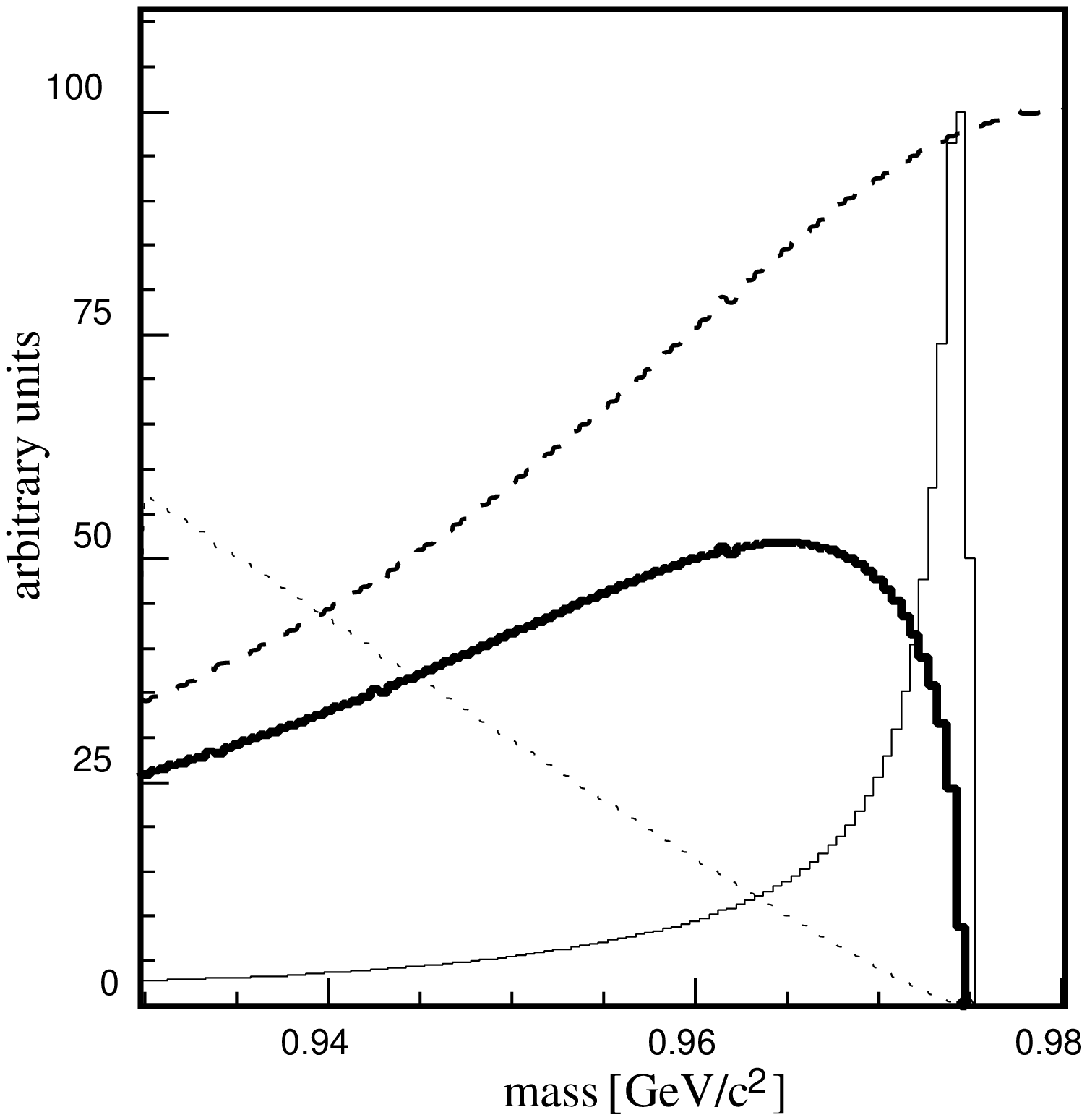,width=0.44\textwidth}}\\

\vspace{-1.cm}

\parbox{0.45\textwidth}{\mbox{}} \parbox{0.05\textwidth}{(a)}
\parbox{0.43\textwidth}{\mbox{}} \parbox{0.05\textwidth}{(b)}\\

\vspace{+0.2cm}

\parbox{1.0\textwidth}{\caption{ \small
           {\bf (a)} Breit-Wigner distribution with
           a mean value of m$_0$~=~980~MeV/$c^{2}$ and a 
           width  $\Gamma$~=~70~MeV/$c^{2}$.
           If the measurement would be performed with the $\sqrt{s}$ as
           depicted
           by the dashed line then the mass m would be measured with
           the
           centre-of-mass excess energy equal~to~Q.\protect\\
           {\bf (b)} {\bf Simulations:}
            Horizontal axis denotes the mass
            of the produced system X via the $pp\rightarrow pp X$ reaction. 
            Long--dashed line denotes a fraction of the Breit-Wigner distribution,
            as shown in figure~\ref{bw1}a.
            Thin--solid line represents the efficiency of the COSY-11
            detection system for the simultaneous detection of the two outgoing
            protons
            from the $pp\rightarrow pp X$ reaction. 
            Short--dashed line shows the decrease of the
            phase space
            volume with the increase of the created mass.
            The phase space  volume was weighted by the proton-proton FSI
            enhancement factor
            taken from reference~\cite{faldt}.
            The thick--solid line  results as a convolution of the above
            described distributions.
           \label{bw1}
        }}
 \end{figure}
Performing the measurement with the total centre-of-mass energy $\sqrt{s}$
close to the sum of the average mass of
the particles in the exit channel,
different mass ranges of the broad particle are populated with
different excess energies: Q(m)=$\sqrt{s} - (2 \cdot m_{p} + m)$
(see figure~\ref{bw1}a).
This implies that the observed mass distribution appears to be different
from the one which would be determined at an excess energy  much larger than
the width of the broad meson.
Neglecting dynamical effects, in case of large $\sqrt{s}$ 
the mass distribution can be roughly approximated by the
Breit-Wigner function, as shown in figure~\ref{bw1}a.

The measurements of the $pp\to ppX$ reactions, reported here, 
have been performed by means of the 
COSY-11 detection system~\cite{COSY11}.
The thick solid line in figure~\ref{bw1}b shows the expected 
missing mass distribution simulated for the total centre-of-mass
energy $\sqrt{s}$ equivalent to a maximal produced mass ($\sqrt{s}~-~2m_{p}$)
smaller than the average mass of the simulated meson.
The shape of this curve results from convoluting the meson spectral density
(long-dashed line) with both (i) the decrease of the phase space volume
with decreasing excess energy Q weighted by the proton-proton FSI enhancement
factor (short--dashed line)
and with (ii) the efficiency of the COSY-11 detection system for the
simultaneous registration of the two outgoing protons from the
$pp\rightarrow pp X$ reaction (thin solid line).\\
The number of observed events per mass bin $\displaystyle\frac{dN}{dm}$ can be written as:
\begin{equation}
  \frac{dN}{dm}(m,Q) = \frac{d\sigma}{dm}(m,Q) \cdot L \cdot E_{ff}(Q),
  \label{eq1}
\end{equation}
with L and E$_{ff}$ denoting the integrated luminosity and the efficiency of the
COSY-11 detection system, respectively.
As proven by extensive Monte-Carlo simulations,
the COSY-11 efficiency is in a good approximation
independent of the produced mass and depends on the
excess energy Q only. The cross section $\displaystyle\frac{d\sigma}{dm}$
 for the creation of a mass m
is expressed by:
\begin{eqnarray}
   \label{eq2}
   \frac{d\sigma}{dm}(m,\ Q) & = &   
      |M(Q)|^{2} \cdot 
      V_{ps}(Q) \cdot FSI_{pp}(Q)
      \cdot SD(m,m_{0},\Gamma) \\
    \frac{d\sigma}{dm}(m,\ Q)
    &  = & \sigma_{primary}(Q) \cdot SD(m,m_{0},\Gamma),
     \label{eq3}
\end{eqnarray}
where $M(Q)$ stands for the matrix element accounting for the production mechanism,
and the last term $SD$ denotes the spectral density of the produced meson
with the average mass and width equal to $m_{0}$ and $\Gamma$, respectively.
The $V_{ps}(Q)$ and $FSI_{pp}(Q)$ represent the phase space volume
available to the outgoing particles and the proton-proton
FSI enhancement factor, respectively~\cite{review}.
The term $\sigma_{primary}$(Q) combines all
excess energy dependent factors
and could be referred to as a primary production cross section 
of producing a mass bin of the resonance which is by a value of Q
smaller than the maximum kinematically available mass.
It should be noted that all factors in equations~\ref{eq1} and~\ref{eq2} depend
on the excess energy Q only, except the spectral density which is a function of
the created mass m.
This means that the number of events
per mass bin can be expressed as follows:
\begin{equation}
     \frac{dN}{dm}(m=\sqrt{s}-2m_{p}-Q,\ Q)  
      = const.(Q) \cdot SD(m,m_{0},\Gamma).
  \label{eq4}
\end{equation}
Scanning the resonance experimentally by changing the value of
$\displaystyle\sqrt{s}=2m_{p}+m+Q$ and keeping in the analysis a
constant value of Q allows to reproduce the mass distribution of
the created meson, even without knowing exactly the energy dependence of the
detection efficiency and other factors. 
Note that if $\sqrt{s} - 2m_{p}$ is within the range of the broad resonance (see
figure~\ref{bw1}a) then
not all parts of the resonance can be created. Thus,
it is not trivial from such a single measurement how to extract the total cross
section. However, by scanning the resonance with different values
of $\sqrt{s} - 2m_{p}$ one can define the total cross section
at a given excess energy Q as an integral over the whole resonance region of 
the  differential cross section $\displaystyle \frac{d\sigma}{dm}(m,\ Q)$
keeping in the integration Q as a constant, thus:
\begin{equation}
 \hspace{-0.9cm}\sigma(Q)=\!\displaystyle\int_{2m_{p}}^{\infty}
 \frac{d\sigma}{dm}(m\!=\!\sqrt{s}-2m_{p}-Q,\ Q) ~d\sqrt{s}
 \label{eq5}
\end{equation}
Please note that if the spectral density function is normalized to unity 
than $\sigma(Q)$ is directly equal to the  primary cross section $\sigma_{primary}(Q)$.
This can be  inferred by substituting equation~\ref{eq3} in equation~\ref{eq5}. \\
Assuming that similar to the production process for pseudoscalar mesons the
value of $|M|^{2}$ will be constant with small values of Q~\cite{review,moskalm0},
one expects that the energy dependence of $\sigma(Q)$
from equation~\ref{eq5} will  be determined by the
Q-dependence of the phase space and the dominant proton-proton FSI.

\section{Measurement of the $pp \to ppX$ reaction close-to-threshold 
         for the production of $f_{0}$ and $a_{0}$ mesons}  
\begin{figure}[H]
\parbox{0.4\textwidth}{\epsfig{file=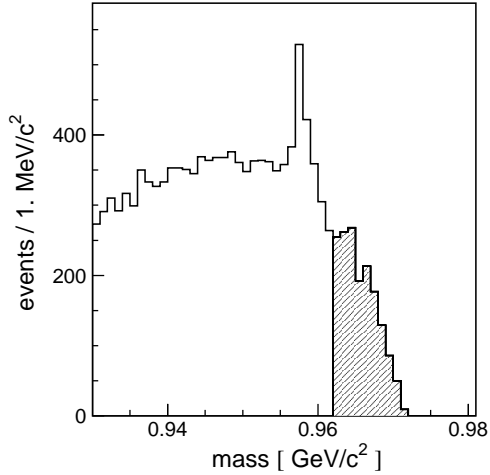,
                        width=0.6\textwidth}}
\parbox{0.6\textwidth}{\caption{\small {\bf Experiment:} 
            Missing mass distribution determined for the $pp\rightarrow ppX$
            reaction at 14~MeV excess energy above the threshold for the
            $\eta^{\prime}$ meson production~\cite{moskalpl}.
            The peak at 958~MeV/c$^{2}$  corrsponds
            to the production of the $\eta^{\prime}$ meson.
            The shaded area denotes the part of the spectrum used in the
            present analysis.
            \label{miss_Q14}
           }}
 \end{figure}

In the following we will present the analysis of the measurements which were
primordially devoted to production studies of the $\eta^{\prime}$ meson in 
proton-proton collisions~\cite{moskalpl}. The experiment was performed at seven
different beam momenta (listed further down in table~\ref{tabelka}).
The  maximum mass ($\sqrt s - 2m_{p}$) available in these experiments covered a
range from 959.6~MeV/c$^2$ to 981.4~MeV/c$^2$. 
This is the region below and close to the average mass
of the f$_0$(980) and a$_0$(980) masses, and hence the signal from the production
of these mesons should indeed influence the overall observed missing mass spectra.
Figure~\ref{miss_Q14} presents one of the missing mass distributions
extracted from the experimental data taken at  a total centre-of-mass energy 
equivalent to the maximum created mass of M = 972~MeV/c$^2$.\\
Certainly the shape of the ``background'' below the clear $\eta^{\prime}$ peak differs
from the shape of the missing mass distribution simulated for the 
Breit-Wigner like resonance 
shown as a thick solid line in figure~\ref{bw1}b. This is due to the fact
that the experimental missing mass spectrum comprises also contributions
from non-resonant multi--pion, $\pi\eta$ and $\pi\omega$  production. 
Thus, taking also into account the well known yield originating
from the $\eta^{\prime}$ production, one can generally extend equation~\ref{eq4}
to:
\begin{equation}
\hspace{-2.5cm}
     \frac{dN}{dm}(m,Q)  =   const.(Q) \cdot
     SD(m,m_{0},\Gamma)  
      +   \frac{dN}{dm}(multi-\pi,\pi\eta,\pi\omega,\sqrt{s})
        + \frac{dN}{dm}(\eta^{\prime},\sqrt{s}) .
  \label{eq6}
\end{equation}
A mass bin corresponding to
Q~=~5~MeV $\pm$ $\Delta$~Q with $\Delta$~Q~=~5~MeV
 was chosen in the analysis for each of the seven
measurements under investigation (see as an example the shaded area in
figure~\ref{miss_Q14}).
The number of events per this mass bin 
has been calculated for seven measurements and after the correction for the
detection efficiency has been normalized to the
corresponding integrated luminosity: 
\begin{eqnarray}
\hspace{-2.5cm}
  \Delta{N}(\sqrt{s}) & = & \frac{1}{L} \displaystyle\int_{Q=0~MeV}^{Q=10~MeV}\frac{dN}{dm}
                            \frac{1}{ E_{ff}(Q)} \ dm \\
  \nonumber
    & = & \displaystyle\int_{Q=0~MeV}^{Q=10~MeV} 
         \sigma_{primary}\cdot \ SD(\sqrt{s}-2m_{p}-Q,m_{0},\Gamma) \ dm +\\
  \nonumber
    & + & \displaystyle\int_{Q=0~MeV}^{Q=10~MeV}
     \left( \frac{d\sigma}{dm}(multi-\pi,\pi\eta,\pi\omega,\sqrt{s})
     + \frac{d\sigma}{dm}(\eta^{\prime},\sqrt{s}) 
     \right) \ dm
\label{eq7}
\end{eqnarray}

\vspace{-1.0cm}

\begin{figure}[H]
\vspace{-0.5cm}
\parbox{0.4\textwidth}{\epsfig{file=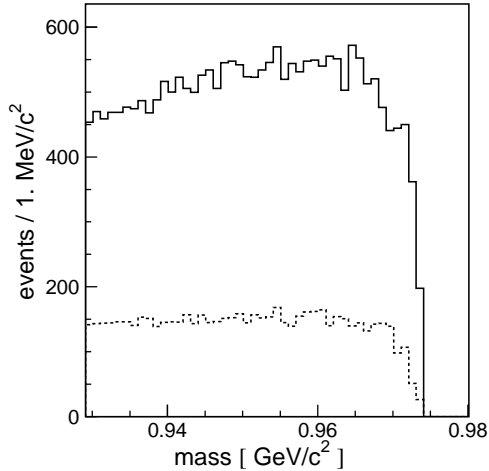,width=0.6\textwidth}}
\parbox{0.6\textwidth}{\caption{ \small {\bf Simulations:}
           Missing mass distributions for the reactions $pp\rightarrow pp2\pi$
           (dashed line) and $pp\rightarrow pp3\pi$ (solid line).
           Please note, that the shape of these spectra differ from each
           other
           and both are different from the missing mass distribution shown
           in figure~\ref{bw1}b. 
           \label{3pi_2pi}
         }}
 \end{figure}
Simulated distributions for the case
of the direct 2$\pi$ and 3$\pi$ creation are shown in figure~\ref{3pi_2pi}.
One can observe that their shapes differ (i) from each other and
(ii) again from the one determined experimentally.
The latter discrepancy can be attributed at least qualitatively to contributions
of the productions of the $f_0(980)$ and/or $a_0(980)$ mesons to the
experimental spectrum.\\

\begin{figure}[H]
\vspace{-0.8cm}
\parbox{0.55\textwidth}{\epsfig{file=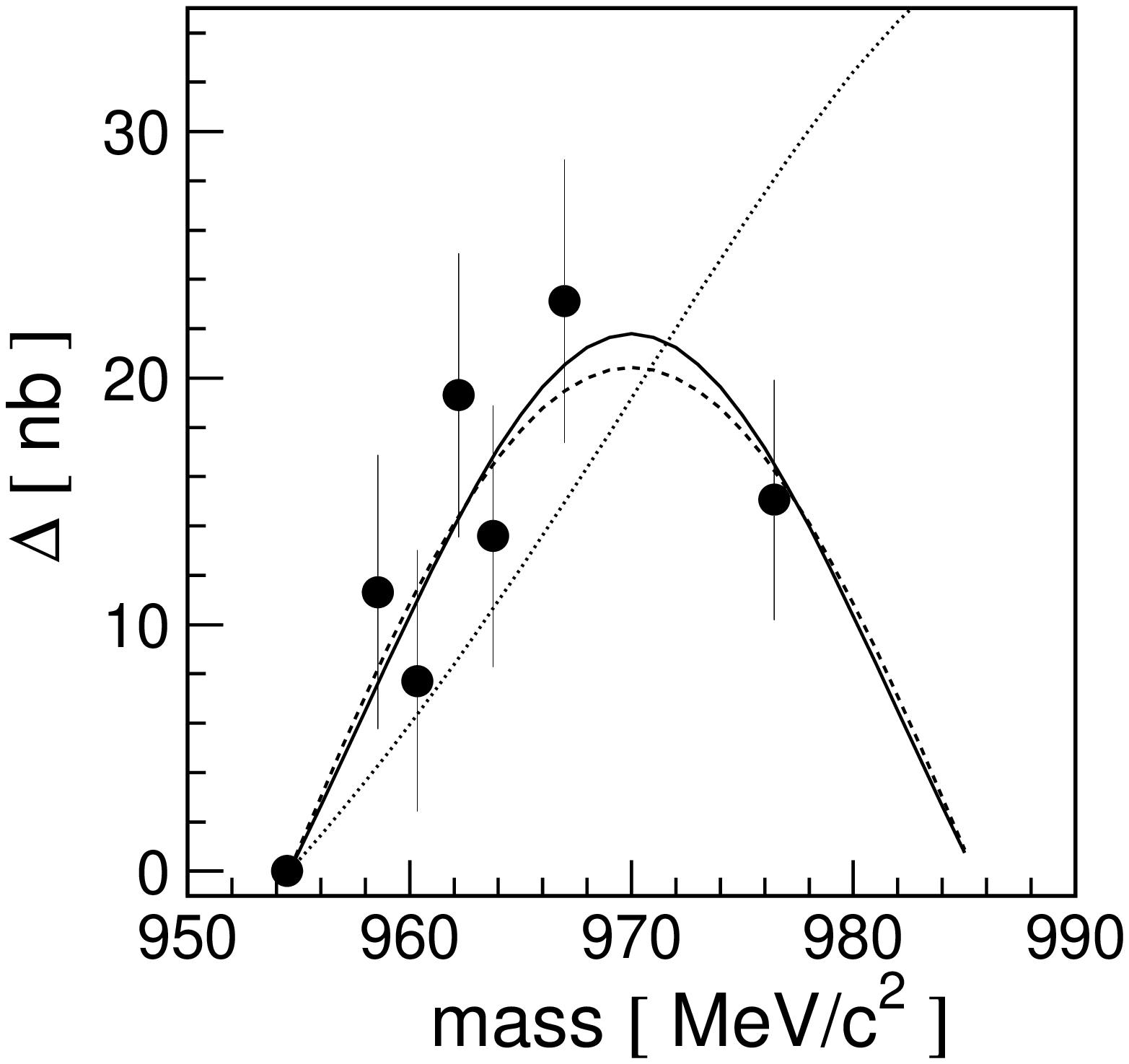,width=0.55\textwidth}}
\parbox{0.45\textwidth}{\epsfig{file=oller.epsi,width=0.45\textwidth}}\\

\vspace{-0.5cm}
\parbox{0.45\textwidth}{\mbox{}}\parbox{0.05\textwidth}{(a)}
\parbox{0.45\textwidth}{\mbox{}}\parbox{0.05\textwidth}{(b)}

\parbox{1.0\textwidth}{\caption{ \small
        {\bf (a)} Points denote number of events, measured  for the 10~MeV upper bin 
          of the missing mass spectrum for the $pp\rightarrow ppX$
          reaction,
          as depicted by the shaded area in figure~\ref{miss_Q14},
          normalized to the integrated luminosity. The value obtained
          at 959.6~MeV/c$^{2}$ was subtracted  from each point. The x-axis
          denotes the  mass corresponding to  Q~=~5~MeV. For example the
          point
          corresponding to the spectrum
          of figure~\ref{miss_Q14} ($\sqrt{s}-2m_{p}$=972~MeV) is plotted
          at a mass value
          of 967~MeV/c$^{2}$.
          Lines show the result of simulations of the $pp\rightarrow ppX$
          reaction, with X being the Breit-Wigner type 
          meson resonance of the following
          parameters:\protect\\
          (dotted line) mass 990~MeV/c$^{2}$, width 65~MeV/c$^{2}$, and $\sigma_{primary}=750$~nb,
          \protect\\
          (solid line) mass 970~MeV/c$^{2}$, width 40~MeV/c$^{2}$, and $\sigma_{primary}=400$~nb,
          \protect\\
          (dashed line) mass 970~MeV/c$^{2}$, width 65~MeV/c$^{2}$, and $\sigma_{primary}=1200$~nb. 
          \protect\\
        {\bf (b)} $\pi^{+}\pi^{-}$ event distribution in the $J/\Psi \to \phi\pi^{+}\pi^{-}$
          decay around the $f_{0}$ mass. Data were taken by DM2~\cite{DM2}
          and MARK-III~\cite{MARK-III} collaborations.
          Solid line depicts the result of the
          calculations of reference~\cite{meissner}.
          \label{f0_cross}
         }}
 \end{figure}

Subsequently, if necessary, the well known values of 
$\displaystyle\frac{d\sigma}{dm}(\eta^{\prime},\sqrt{s})$~\cite{moskalpl}
were subtracted from the above defined $\Delta{N}(\sqrt{s})$:
\begin{equation}
 \Delta{N^{\prime}}(\sqrt{s}) = \Delta{N}(\sqrt{s}) - \displaystyle\int_{Q=0~MeV}^{Q=10~MeV}
  \frac{d\sigma}{dm}(\eta^{\prime},\sqrt{s}) \ dm
  \label{eq8}
\end{equation}
Further,  in order to account for the contribution
from the multi-pion, $\pi\eta$ and $\pi\omega$ 
production we subtracted from each 
$\displaystyle \Delta{N^{\prime}}(\sqrt{s})$
the value determined for the lowest measured energy.
The result  is shown in figure~\ref{f0_cross}a, where the variable $\Delta$ 
at the vertical axis reads:
\begin{equation}
 \Delta =  \Delta{N^{\prime}}(\sqrt{s}) 
 - \Delta{N^{\prime}}(\sqrt{s}_{lowest}~=~2 \cdot m_p + 959.6~MeV)
\label{eq9}
\end{equation}
On the other hand,
assuming changes of the cross section for the non-resonant multi-pion
$\pi\eta$ and $\pi\omega$ production
to be negligible 
the value of
$\Delta $ can be expressed as follows:
\begin{eqnarray}
 \nonumber
 \Delta & = & \sigma_{primary} \cdot \displaystyle\int_{Q=0~MeV}^{Q=10~MeV}
          SD(m,\sqrt{s}-2m_{p}-Q,\Gamma) \ dm  \\
        & - &\sigma_{primary} \cdot  \displaystyle\int_{Q=0~MeV}^{Q=10~MeV}
          SD(m,\sqrt{s}_{lowest}-2m_{p}-Q,\Gamma) \ dm
  \label{eq10}
\end{eqnarray}
Values determined for $\Delta$ are given in table~\ref{tabelka} and also 
shown in figure~\ref{f0_cross}a.

The curves in this figure 
represent calculations  performed according to equation~\ref{eq10}
approximating the spectral function  SD$(m,m_{0},\Gamma)$ by the Breit-Wigner distribution.
The obtained result
demonstrates that
the data are sensitive to the average mass of the created meson
or mesons but rather non-sensitive to the width of an assumed Breit-Wigner structure.
The latest volume of the Review of Particle Physics~\cite{PDG} shows that the
parameters for $f_0(980)$ and $a_0(980)$ mesons are very similar to
each
other:\\
$f_0(980): ~~m = 980 \pm 10$ MeV,~~~~~~~$\Gamma = 40$~to~100~MeV \\
$a_0(980): ~~m = 984.7 \pm 1.2$ MeV,~~~~$\Gamma = 50$~to~100~MeV \\
and therefore we cannot distiguish between these two resonances in a missing mass
analysis from the proton--proton interaction. 

\vspace{-1cm}
\hspace{-2.0cm}
\parbox{0.5\textwidth}{
\begin{figure}[H]
\vspace{+0.5cm}
\hspace{-1cm}
{\caption{\small
         $\chi^{2}$ as a function of the total cross section.
         Number of degrees of freedom is equal to 3.
         \label{chi2vssigma} 
        }}
\vspace{-0.8cm}
\hspace{1.5cm}
{\epsfig{file=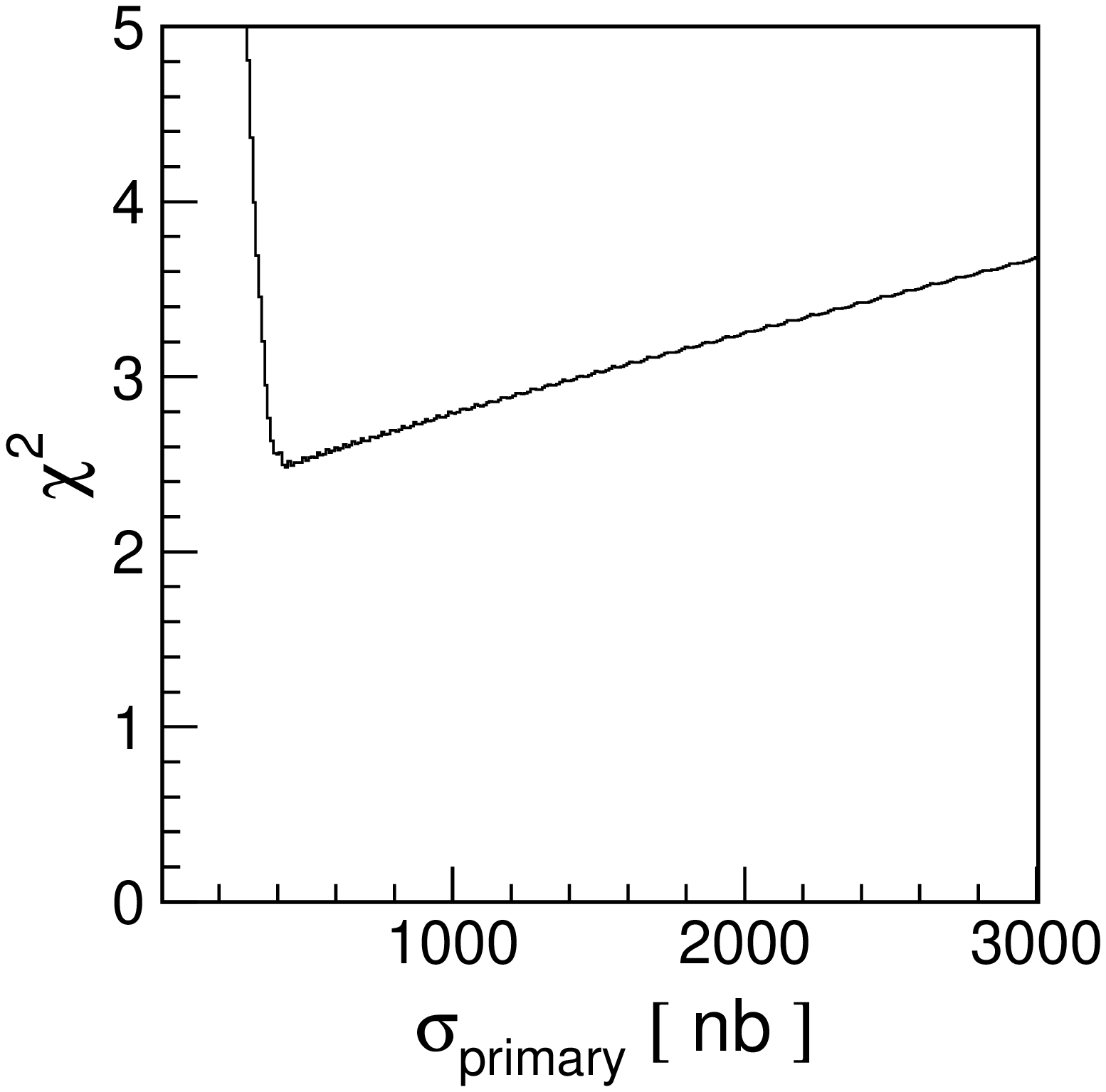,width=0.5\textwidth}}
\end{figure}
}
\hspace{-1.3cm}
\parbox{0.5\textwidth}{
\vspace{-0.6cm}
\begin{table}[H]
\caption{\label{tabelka} 
Result of the analysis of the $pp \to pp X$ reaction measured at COSY-11;
$\Delta$ is defined in equation~\ref{eq10} }.
\begin{indented}
\item[]\begin{tabular}{@{}llll}
\br
Beam          & maximum     & $\Delta$ & $error(\Delta)$ \\

momentum      & mass of X   &  & \\
  MeV/c       &  MeV/ c$^2$ &  nb  &  nb \\
\mr
3214.0 &  959.6   &   0.0   &   0.0 \\
3226.6 &  963.6   &  11.3   &   5.6 \\
3232.1 &  965.4   &   7.7   &   5.3 \\
3238.0 &  967.2   &  19.3   &   5.8 \\
3242.9 &  968.8   &  13.6   &   5.3 \\
3253.1 &  972.0   &  23.1   &   5.7 \\
3282.9 &  981.4   &  15.1   &   4.9 \\
\br
\end{tabular}
\end{indented}
\end{table}
}

 A fit
 to the resonance--like structure of figure~\ref{f0_cross}a results in values of:\\
 $m_{0} = 969^{+3}_{-2}$~MeV       
 and $\sigma_{primary}$~=~430$^{+2070}_{-100}$~nb
 and is nearly identical to the solid line in figure~\ref{f0_cross}a.
 The quality of the fit can be estimated from the $\chi^2$ distribution
 shown in figure~\ref{chi2vssigma}.
 It is worth to note, that the obtained structure is in line
 with the measurements of the $\pi^{+}\pi^{-}$ invariant mass distribution
 of the $J/\Psi$ decay into $\phi\pi^{+}\pi^{-}$ system as shown in Figure~\ref{f0_cross}b.
 It is rather well justified that the 
 the total cross section of multi-pion and $\pi\eta$  changes only insignficantly 
 over the studied 22~MeV excess energy range since the present measurements were performed
 some 560~MeV and 300~MeV above the 3$\pi$ and $\pi\eta$ thresholds. 
 Still, it cannot be excluded that the variations of $\Delta$ 
 are due to the growth of an unknown cross section for $\pi\omega$ production 
 since in this case the excess energy change from 42 to 64~MeV.
 Therefore the determined value of the total 
 cross sections for $a_{0}$/$f_{0}$ resonances
 can only be treated as an upper limit.

\section{Conclusion}
For the example of the production of the f$_0$(980) and/or a$_0$(980) mesons
we discuss the notion of the reaction threshold in case of broad resonances
where the available phase space volume for a given mass bin changes
significantly.
Ansatz of equation~\ref{eq2} leads to a simple definition
of the differential~- and total cross section also in the case of the close-to-threshold
production of a broad resonance.

The performed measurements of the $pp\to ppX$ reaction at seven beam momenta
close to the threshold of  $f_0$ and $a_0$ mesons
revealed an enhancement in the missing mass spectrum.
If the observed structure were due to the production 
of the object possessing a  Breit-Wigner form
its total cross section introduced in equation~\ref{eq5}
would be equal to about 430~nb at excess energy of Q~=~5~MeV.
 However, due to the unknow influence of the
$\pi\omega$ production we can treat this value only as an estimation 
of an upper limit for the production of $f_0$ and $a_0$ meson via 
the $pp\to pp X$ reaction. Here we would like to stress that 
this is the first experimental investigation
utilizing a missing mass techniques
of the production of these scalar resonances close to the kinematical threshold.
 The determined upper limit of the total cross section 
is much larger than expected from  existing
theoretical estimations, which however take into account only 
a part of the possible production  mechanisms. 
In case of $f_0$, for example, one could consider also 
a production via an excitation of  baryonic resonances.
PDG~\cite{PDG} reports a few $N^*$ resonances around 1700 MeV with widths
of 50-250 MeV, which decay predominantly into the $N\pi\pi$ channel.

\vspace{0.3cm}
{\bf Acknowledgements:}\\
  We would like to thank  J.~A. Oller for comments
  to the first version of the manuscript. \\
 The work has been partly supported by the European Community - Access to
Research Infrastructure action of the Improving Human Potential Programme
as well as the International B\"uro and the Verbundforschung of the BMBF.


\end{document}